\documentclass{article}

\usepackage{arxiv}

\usepackage[title]{appendix}
\usepackage[utf8]{inputenc} 
\usepackage[T1]{fontenc}    
\usepackage[numbers]{natbib}
\usepackage{booktabs}       
\usepackage{amsfonts}       
\usepackage{amsmath,bm,upgreek}
\usepackage{hyperref}       
\usepackage{url}            
\usepackage{nicefrac}       
\usepackage{microtype}      
\usepackage{cleveref}       
\usepackage{graphicx}
\usepackage{doi}
\usepackage{subfig}

\graphicspath{ {./images/}}

\DeclareMathOperator{\T}{\mathsf{T}}
\DeclareMathOperator{\sech}{sech}

\title{Reciprocal hydrodynamic response estimation in a random spreading sea}

\date{June, 2023}

\author{ \href{https://orcid.org/0000-0001-8323-7406}{\includegraphics[scale=0.06]{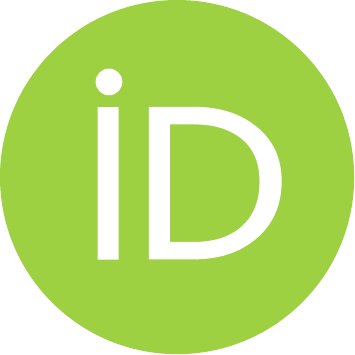}\hspace{1mm}Jiannan Yang}      
   \thanks{corresponding author: \href{mailto:jy419@cam.ac.uk}{jy419@cam.ac.uk}} 
   \qquad
    Robin Langley
    \qquad
    Richard Lines \\
   \\
  	Department of Engineering\\
	University of Cambridge\\
	Trumpington Street, Cambridge 
	CB2 1PZ, UK\\
}

\renewcommand{\shorttitle}{\textit{DFR}}

\hypersetup{
pdftitle={},
pdfsubject={},
pdfauthor={Jiannan Yang},
pdfkeywords={},
}

\begin{document}
\maketitle

\begin{abstract}
Direct estimation of the hydrodynamic response of an offshore structure in a random spreading sea can lead to large computational costs. In this paper the actual spreading sea is replaced by an idealised diffuse wave field and the diffuse field reciprocity (DFR) relationship is derived analytically and verified against diffraction analysis for offshore application.  The DFR approach provides an analytical expression for the estimation of the wave loading spectrum in a spreading sea. It is very efficient because only the added damping coefficients are required. Furthermore, if normalised to the peak amplitude of a spreading sea, an upper bound response can be obtained using the reciprocal approach. And this is demonstrated using a spar type floating wind turbine. Given that the hydrodynamic coefficients are routine outputs for offshore structural design, engineers would obtain the upper bound response without additional computational cost using this new approach. 
\end{abstract}


\section{Introduction}
\label{sec:1}
Realistic sea states contain waves that are dependent on frequency as well as direction. The response of an offshore structure can be sensitive to both the frequency and the heading of an incident ocean wave, and the most realistic way of ensuring that this effect is captured in a numerical simulation is to model the wave environment as a random spreading sea. 

For example, large floating offshore structures subject to multi-directional random waves have been studied in \cite{Isaacson1986Directional}, where the importance of accounting for directional spreading in wave force and motion response calculation is highlighted. In \cite{Wei2016Directional}, studies of a four leg jacket supported offshore wind turbine at three different sites show that the structural capacity is sensitive to the load directions, and a unidirectional sea cannot account for the directional dependence of capacity and loading that will influence the reliability of the offshore system. 

Despite its importance, offshore designers and engineers often adopt a unidirectional sea for their hydrodynamic analysis. This is partly due to the fact that measuring directionality is more complicated than measuring the free surface elevations \cite{McAllister2017Wave, Adcock2009Estimating}, also because accounting for a spreading sea can lead to large computational costs. For example, if a frequency domain analysis is performed then the dynamic response must be computed at a large number of combinations of frequency and wave headings, and this can lead to large computational costs, particularly so at the early design stage when many possible design configurations need to be analysed. The motivation of this paper is thus to look for an efficient approach to estimate the hydrodynamic response in a spreading sea. 

In this paper, to reduce the computational cost, the actual spreading sea is replaced by an idealized diffuse wave field, and a technique from the vibro-acoustic literature known as “diffuse field reciprocity (DFR)” \cite{Shorter2005On} is used to express the cross spectrum of the wave loading in terms of the potential damping matrix of the structure. 

In room acoustics, a diffuse sound filed is an ideal probabilistic model describing a sound field consisting of a very large set of statistically uncorrelated elemental plane waves. The propagation direction of the waves are random with a uniform probability distribution \cite{Fahy2005Foundations}. This enables the wave field, instead of the individual waves, to be treated in a stochastic manner. In analogy to the fluid media, the diffuse field concept has also been applied to waves in structures \cite{Egle1981Diffuse, Evans1999Measurement}. Furthermore, the wave approach to formulate the widely used Statistical Energy Analysis (SEA) is mainly based on the diffuse field assumption \cite{LafontT2014Review}. 

An efficient approach to tackle this vibroacoustic problem has been achieved by using the diffuse field reciprocity (DFR) principle \cite{Shorter2005On}. This principle states that the loading applied by a diffuse wave field on a deterministic structure can be expressed in terms of the energy in the diffuse field and the radiation properties of the structure into an infinite space. This methodology has been applied to numerous systems in vibro-acoustics. In addition, it is shown in \cite{Langley2010Reciprocity} that the diffuse field reciprocity principle can also be applied to electromagnetic systems, enabling the currents induced in a wiring system by diffuse electromagnetic waves to be computed in an efficient manner. 

It is shown in this paper that the hydrodynamic response of an offshore structure can be successfully recovered through the reciprocal DFR principle in an idealised diffuse sea environment. In this way, instead of solving a full diffraction problem, only the radiation potential in still water needs to be computed. In addition, it is possible to estimate an upper bound response in a spreading sea assuming the worst sea state is coming from all directions. Compared to the commonly used unidirectional sea assumption, the DFR approach provides a new reciprocal approach that is as efficient, considers all degree of freedoms and provides a higher safety factor. This approach is weakly analogous to the Haskind relation \cite{Newman2018Marine} that is employed in diffraction theory, although in the present case the cross spectra of the wave loading are considered rather than the diffraction forces. 

In what follows, the formulation of the diffuse field reciprocity (DFR) principle is introduced in Section \ref{sec:2}. The analytical verification of the proposed reciprocal approach is given in Section \ref{sec:3}, in terms of the hydrodynamic response of a simple articulated buoy, by demonstrating the equivalence between the reciprocal DFR approach and the direct approach via a full diffraction analysis. In Section \ref{sec:4}, an upper bound is introduced together with discussions of how the upper bound can be estimated efficiently using the DFR approach. Following the discussions, a demonstration case study using a spar type floating wind turbine is presented in Section \ref{sec:5}. Concluding remarks are given in Section \ref{sec:6}. 
\section{Hydrodynamic response in a spreading sea }
\label{sec:2}
For a linear system subject to a random wave excitation, the covariance matrix of the structure response at degrees of freedom $\zeta$ can be found as \cite{DavidENewland1994introduction}:
\begin{equation} \label{eq:1}
      \bm{\sigma}^2_\zeta = \iint \mathbf{G} (\omega,\theta)\mathbf{G}^{\mathsf{^{*}T}}(\omega,\theta)S_{\eta\eta} (\omega,\theta) d\omega d \theta
\end{equation}
where $\mathbf{G}(\omega,\theta) \in \mathbb{C}^{\mu}$ is the transfer function, with $G_j$  relates the structural response at the degree-of-freedom $\zeta_j$ to a surface wave of unit elevation. $S_{\eta\eta}(\omega,\theta)$ is the directional spectrum of the surface wave elevation at frequency $\omega$ with wave direction $\theta$. The notation $[\cdot]^{\mathsf{^{*}T}}$ indicates complex conjugate transpose. 

A blocked force is the force experienced by the structure when it is subject to incoming waves but held stationary \cite{yang2015non, bobrovnitskii2001theorem}. Using this concept, the transfer function $G(\omega,\theta)$ can be further defined as:
\begin{equation} \label{eq:2}
      \mathbf{G} (\omega,\theta) = \mathbf{H} (\omega)\mathbf{f}_b (\omega,\theta)
\end{equation}
where $\mathbf{f}_b$ is the blocked force due to a wave of unit elevation. Note that the blocked force is dependent both on frequency and the incoming wave direction.  

$\mathbf{H} \in \mathbb{C}^{\mu \times \mu}$ is the structural transfer function matrix, which for a linear system can be found as the inverse of the dynamic stiffness matrix:
\begin{equation} \label{eq:3}
      \mathbf{H} (\omega) = \left[  -\omega^2\mathbf{M}  + i\omega\mathbf{C} + \mathbf{K} \right]^{-1}
\end{equation}
where $\mathbf M$, $\mathbf C$ and $\mathbf K$ are the inertia, damping and stiffness matrices. 

For offshore engineering, it is a common practice to consider a frequency independent spreading function $D(\theta)$.  The directional spectrum of the wave elevation can then be written as \cite{SarpkayaT1981Mechanics}:
\begin{equation} \label{eq:4}
      S_{\eta\eta}(\omega,\theta) = S_{\eta\eta}(\omega)D(\theta)
\end{equation}
where $S_{\eta\eta}(\omega)$ is the unidirectional wave spectrum, and the following condition must be satisfied for the spreading function:  
\begin{equation} \label{eq:5}
      \int_0^{2\pi} D(\theta) d\theta = 1
\end{equation}
Therefore, Eq \ref{eq:1} can be expanded as:
\begin{equation} \label{eq:6}
      \bm{\sigma}^2_\zeta = \int \mathbf{H} (\omega) 
      \left[ \int_0^{2\pi} \mathbf{f}_b (\omega,\theta) \mathbf{f}_b^{\mathsf{^{*}T}}(\omega,\theta) D(\theta) d \theta \right] 
     \mathbf{H}^{\mathsf{^{*}T}}(\omega)S_{\eta\eta} (\omega,\theta) d\omega
\end{equation}
Eq \ref{eq:6} indicates that a two dimensional integration over wave heading as well as frequency are required for the estimation of the hydrodynamic response in a spreading sea. As a large number of wave headings is normally required (e.g., as much as 12 headings are recommended in \cite{DNV2002Recommended}), the computational cost for analysis in a spreading sea can be prohibitive, especially at early design stage. Motivated to overcome this issue, we focus on simplifying the part of integration that is taken over the wave headings $\theta$ (the expression inside the square brackets of Eq \ref{eq:6}). 

The key insight here is that the spreading function $D(\theta)$ in Eq \ref{eq:5} can be seen as a measure for the fraction of sea states, out of a large number of ensemble, that are coming from direction $\theta$. In other words, the wave direction $\theta$ is regarded as a random variable, with its probability distribution described by the density function $D(\theta)$. 

With this view, the expression inside the square brackets in Eq \ref{eq:6} is the expected cross spectrum of the blocked force by ensemble average. The covariance matrix of the hydrodynamic response can then be more compactly expressed as: 
\begin{equation} \label{eq:7}
      \bm{\sigma}^2_\zeta = \int \mathbf{H} (\omega) 
      \mathbb{E}_\theta \left[ \mathbf{f}_b \mathbf{f}_b^{\mathsf{^{*}T}}\right] 
     \mathbf{H}^{\mathsf{^{*}T}}(\omega)S_{\eta\eta} (\omega,\theta) d\omega
\end{equation}
where the expectation is taken over the incident wave headings $\theta$ over 0 to $2\pi$. 

At first sight, it might appear that we haven't gained any computational saving by converting the expression from Eq \ref{eq:6} to Eq \ref{eq:7} by taking the ensemble average point of view, because large computational efforts are still required to compute the expected cross spectrum for the wave forces. 

However, if the spreading sea is replaced by an idealised diffuse wave field, the technique from the vibro-acoustics literature known as “diffuse field reciprocity (DFR)” \cite{Shorter2005On} can be used to express the cross spectrum of the wave loading in terms of the resistive impedance to radiation in a reciprocal manner. In this way, instead of solving a full diffraction problem, only the radiation potential in still water needs to be computed. 

From \cite{Shorter2005On}, the diffuse field reciprocity (DFR) relationship is given as: 
\begin{equation} \label{eq:8}
      \mathbb{E} \left[ \mathbf{f}_b \mathbf{f}_b^{\mathsf{^{*}T}}\right]  = 
     \frac{4E(\omega)d\omega}{\pi\omega n(\omega)} \text{Im} (\mathbf{D}_\text{dir}(\omega))
\end{equation}
where $\mathbb{E}[\cdot]$ is the mathematical expectation or ensemble average and $\mathbf{f}_b$ represents the vector of blocked forces. On the right hand side of Eq \ref{eq:8}, $E(\omega)$ and $n(\omega)$ are the spatially averaged system energy and modal density respectively \cite{Woodhouse1981introduction}, and are both frequency dependent. The matrix $\mathbf{D}_{\text{dir}}$ is the dynamic stiffness matrix describing the direct filed radiation. 

The diffuse field reciprocity (DFR) from Eq \ref{eq:8} relates the cross spectrum of the blocked force to the resistive part of the direct field dynamic stiffness $\mathbf{D}_{\text{dir}}$. This corresponds to the wave energy radiated into the far field. For a body moving on or near a free water surface, energy got carried away from the moving structure by travelling waves. This results an energy loss and this is called added damping or potential damping $\mathbf{C}_{\text{pot}}$. In addition to the travelling waves, there are also evanescent waves near the structure. They do not dissipate energy and are related to the reactive part of direct field dynamic stiffness matrix $\mathbf{D}_{\text{dir}}$. Therefore, the radiated force due to the structure oscillation in still water can be obtained as:
\begin{equation} \label{eq:9}
       \mathbf{f}_{\text{rad}} = \mathbf{D}_{\text{dir}} \bm{\zeta} =
       (-\omega^2 \mathbf{M}_a + i \omega \mathbf{C}_{\text{pot}}) \bm{\zeta}
\end{equation}
where $\mathbf{f}_{\text{rad}}$ is the force vector due to structure movement and $ \mathbf{M}_a$ is the added mass matrix. Therefore, the imaginary part of $\mathbf{D}_{\text{dir}}$, which is required for the reciprocal relationship in Eq \ref{eq:8}, is directly related to the potential damping matrix, i.e., $\text{Im}(\mathbf{D}_{\text{dir}}) = \omega \mathbf{C}_{\text{pot}}$. 

It can be seen from Eq \ref{eq:8} that the direct integration over the wave headings is reduced to a simple algebraic expression, by taking the ensemble average in a diffuse sea. This allows an efficient estimate for the hydrodynamic response in a spreading sea environment. 

The diffuse field reciprocity (DFR) relation were originally developed in vibro-acoustic field and as a result the parameters in Eq \ref{eq:8} in offshore context are yet to be defined. In Appendix \ref{appendix:a}, the modal density and spatial averaged energy are formulated in an offshore context. Putting all the components together, the DFR relationship in offshore application can be found as:
\begin{equation} \label{eq:10}
    \begin{split} 
      \mathbb{E} \left[ \mathbf{f}_b \mathbf{f}_b^{\mathsf{^{*}T}}\right]  & = 
     \frac{4E(\omega)d\omega}{\pi\omega n(\omega)} \text{Im} (\mathbf{D}_\text{dir}(\omega))  \\
     & =  \frac{2\rho g^2 S_{\eta \eta (\omega)} d\omega}{\omega k} 
     \left[ \tanh (kd) + kd \sech^2(kd) \right] \mathbf{C}_{\text{pot}}
    \end{split}
\end{equation}
where $k$ is the wavenumber, $d$ is the water depth. Note that the above expression is applicable for water of finite depth. For deep water, $kd$ is large, and $\tanh(kd) \approx 1$, $\sech(kd) \approx 0$. Therefore, the DFR relationship in Eq \ref{eq:10} become much simplified. 

The reciprocal DFR relationship in Eq \ref{eq:10} might look unfamiliar to researchers and practitioners in offshore or marine engineering. Nevertheless, relating the exciting forces and moments in terms of the far-field behaviour of the radiation potential is not new and is known as Haskind relation in diffraction theory \cite{Newman2018Marine, Newman1963Exciting}. Except that in the present case, cross spectrum of the wave loading is considered rather than only the diffraction forces.

The cross spectrum of the blocked force of an offshore structure in a diffuse sea can then be obtained analytically once the potential damping matrix is known. In Section \ref{sec:3}, for the purpose of analytical verification, the potential damping coefficients are obtained by computing the radiation force with a diffraction analysis. However, as to be discussed in Section \ref{sec:4.1}, the hydrodynamic coefficients, such as added mass and potential damping, are well documented for typical offshore structures. For new designs, a hydrodynamic model for the hydrodynamic coefficients is usually developed as part of the design procedure \cite{DNV2002Recommended}. Therefore, hydrodynamic analysis using the proposed diffuse field reciprocity approach can take full advantage of the readily available damping coefficients. And this is demonstrated in Section \ref{sec:5} with an example using a spar type floating wind turbine. 

\section{Analytical verification using a simple articulated buoy}
\label{sec:3}
To verify the applicability of the diffuse field reciprocity (DFR) on offshore structures in a diffuse sea environment, we compare the blocked force spectrum calculated from two different approaches: a direct approach, where a diffraction analysis is conducted, and the reciprocal approach using DFR from Eq \ref{eq:10}. 

As stated in the introduction, the hydrodynamic coefficients required for the DFR are mostly well documented or ready to be extracted from standard numerical methods \cite{Gao2011Hydroelastic}. However, for completeness, here we demonstrate briefly the steps to obtain the added mass and damping coefficients by solving the radiation problem analytically for a simple articulated buoy. 

\begin{figure}[!h]
	\centering
	 \subfloat[\centering Side View] {{\includegraphics[width=7cm]{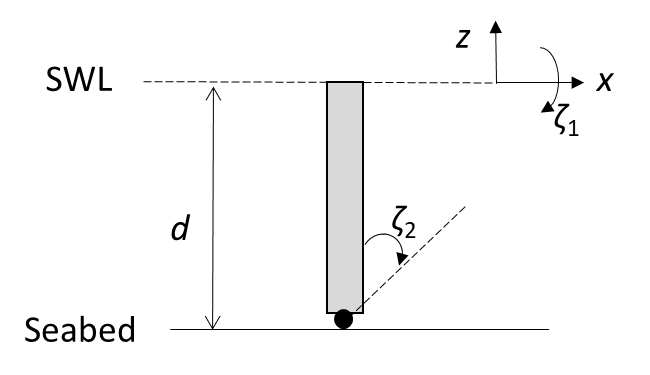} } \label{fig:1a}} 
	 \qquad
    \subfloat[\centering Top View]{{\includegraphics[width=5cm]{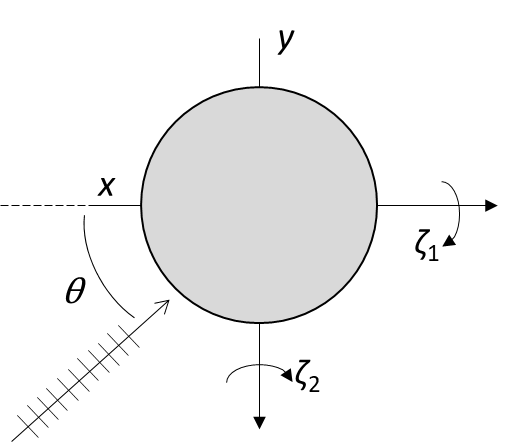} }\label{fig:1b}}
	\caption{A rigid articulated buoy with 2 degrees of freedom subject to a wave incident at angle $\theta$ (SWL: still water line).}
	\label{fig:1}
\end{figure}

The buoy of length $d$, as shown in Figure \ref{fig:1}, is a long and hollow column anchored to the seabed via a ball joint. It has two degrees of freedom, rotation around the $x$ axis (out of plane), and rotation around the $y$ axis (in plane). In this example, the buoy is considered to be a rigid structure with relative large dimensions compared to the wavelength of surface waves so that drag force can be ignored here.
\subsection{\textit{Reciprocal approach} with hydrodynamic coefficients}
\label{sec:3.1}
Using Bernoulli’s equation \cite{SarpkayaT1981Mechanics}, the dynamic pressure acting on a structure can be found via the radiated velocity potential:
\begin{equation} \label{eq:11}
       p_r = - \rho \left( \frac{ \partial \phi}{\partial t} \right)_{a}
\end{equation}
where $a$ is the radius of the buoy and $\phi$ is the velocity potential. Integrating the pressure along the buoy surface, the overall moment acting on the structure can be obtained as: 
\begin{equation} \label{eq:12}
      f_{\text{rad}} = - \int_{-d}^0 \int_0^{2\pi} 
                                 (z+d)p_r \mathbf{n} d\theta dz                           
\end{equation}
where $\theta$ is the circumferential coordinate and $\mathbf{n}$ is the directional normal vector, which is equal to $a\cos\theta$. 

We call $f_\text{rad}$ the radiation moment and it is calculated with respect to the bottom joint of the buoy. Assuming time harmonic dependence, and use the velocity potential $\phi$ for the simple articulated buoy as given in Appendix \ref{appendix:b}, the radiation moment can be found as:
\begin{equation} \label{eq:13}
      f_{\text{rad}} = -\zeta \omega^2 e^{-i \omega t}
         \sum_{p=0}^{\infty}  \frac{\pi \rho a}{k_p}
         \frac{H^{(1)}_1 (k_p a)}{ \left[ H^{(1)}_1 (k_p a) \right]'}
         \frac{\left[  \frac{d}{k_p} \sinh(k_p d) - \frac{1}{k_p^2} (\cosh(k_p d) - 1)  \right]^2} {\frac{1}{4 k_p} \sinh (2 k_p d) + \frac{d}{2}}                      
\end{equation}
where $H^{(1)}_q$ is the Hankel function of first kind and $qth$ order. 

$k_p$ is the wave number and can be found from the dispersion relation, $k\tanh(kd) = \omega^2/g$, via numerical iterative routines such as Newton Raphson’s method. There are two real roots and infinite number of imaginary roots for wave number $k$ at each frequency $\omega$. The two real roots represent waves propagating in opposite directions, while the imaginary values denote the evanescent waves. For the ease of notation, from this point on, we will use $k$ to represent the wavenumber and drop the summation sign. 
 
Given a unit displacement input ($\zeta = 1$), the radiation moment can be found and the potential damping matrix $\mathbf{C}_{\text{pot}}$ can then be obtained as the imaginary part of $f_{\text{rad}}$ from Eq \ref{eq:9} :
\begin{equation} \label{eq:14}
     \mathbf{C}_{\text{pot}} =   \frac{ \text{Im} \left[  f_{\text{rad}} (\zeta = 1) \right] }{\omega}
     = \frac{-4\omega^2 \rho}{k^5 (J_1^{'2}(ka) + Y_1^{'2}(ka) )} 
     \frac{(kd \sinh (kd) + 1 - \cosh (kd))^2}{(\tanh(kd) + kd \sech^2(kd))\cosh^2(kd)}
     \begin{bmatrix}
      1& 0 \\
      0 & 1
    \end{bmatrix}                        
\end{equation}
where $J$ and $Y$ are the Bessel function of first and second kind respectively. The $2 \times 2$ matrix corresponds to the two degrees of freedom of the buoy. As the two rotation motions, $\zeta_1$ and $\zeta_2$ as seen in Figure \ref{fig:1}, are independent due to symmetry, the matrix is diagonal. 

Using the DFR relation in Eq (9), the ensemble average of the cross spectra of the blocked moment can be found as:
\begin{equation} \label{eq:15}
        \mathbb{E} \left[ \mathbf{f}_b \mathbf{f}_b^{\mathsf{^{*}T}}\right]_{\text{reciprocal}}  = 
        S_{\eta \eta} (\omega) d\omega \frac{8\rho^2 g^2}{k^6}
        \left(  \frac{kd \sinh (kd) + 1 - \cosh (kd)}{\cosh(kd)} \right)^2
        \frac{1}{ (J_1^{'2}(ka) + Y_1^{'2}(ka) )} 
     \begin{bmatrix}
      1& 0 \\
      0 & 1
    \end{bmatrix}                        
\end{equation}
Therefore, Eq \ref{eq:15} is the result obtained via the reciprocal approach by using the potential damping matrix and the DFR relationship from Eq \ref{eq:9}. In comparison, the cross spectrum of blocked moments is to be estimated explicitly in the next section via the direct approach using diffraction analysis. 
\subsection{Analytical verification via a \textit{direct approach}}
\label{sec:3.2}
The diffuse field reciprocity (DFR) approach, as described in Eq \ref{eq:10} for offshore applications, relates the expected cross spectrum of the blocked forces to the wave energy radiation into far field in a reciprocal manner. To verify the DFR results, using the articulated buoy subject as an example, the cross spectrum of the blocked force is also derived directly via analytical diffraction analysis. 

For a buoy subject to a single incident wave at angle $\theta$, as shown in Figure \ref{fig:1}, the blocked moment can be found as ([36], [15]):
\begin{equation} \label{eq:3a1}
        f_b = 
        \frac{\pi D^2}{8k}\rho g H
        \left(  \frac{kd \sinh (kd) + 1 - \cosh (kd)}{\cosh(kd)} \right)^2
        C_m \cos(\omega t - \delta)
\end{equation}
where $H$ is wave height.  $C_m$ is the effective inertia coefficient and $\delta$ is the phase angle: 
\begin{align*} 
        C_m = \frac {4 \left[ J_1^{'2}(ka) + Y_1^{'2}(ka) \right]^{-1/2} }{\pi(ka)^2} \qquad  \text{and}  \qquad
        \delta = - \tan^{-1} \left[ \frac{ Y_1^{'2}(ka)} {J_1^{'2}(ka)} \right]
\end{align*}
In the present case for the simple buoy with two degrees of freedom, the blocked moment vector can be expressed simply as: 
\begin{equation} \label{eq:3a2}
        \mathbf{f}_b = f_b
     \begin{bmatrix}
      \cos\theta  \\
      \sin\theta 
    \end{bmatrix}                        
\end{equation}
For excitation due to a random wave, the wave height is then related to the random wave spectrum, i.e., $H = 2\sqrt{2S_{\eta \eta}(\omega) d\omega}$. 

The cross spectrum of the blocked moment can then be obtained as: 
\begin{equation} \label{eq:16}
        \mathbf{f}_b \mathbf{f}_b^{\mathsf{^{*}T}} = 
        S_{\eta \eta} (\omega) d\omega \frac{16\rho^2 g^2}{k^6}
        \left(  \frac{kd \sinh (kd) + 1 - \cosh (kd)}{\cosh(kd)} \right)^2
        \frac{1}{ (J_1^{'2}(ka) + Y_1^{'2}(ka) )} 
     \begin{bmatrix}
      \cos^2\theta & \cos \theta \sin \theta \\
      \cos \theta \sin \theta & \sin^2\theta
    \end{bmatrix}                        
\end{equation}
Eq \ref{eq:16} gives the blocked moment cross spectrum matrix for a single incident wave at angle $\theta$. It follows that the expected cross spectrum of the blocked moment in a directional sea is: 
\begin{equation} \label{eq:17}
       \mathbb{E} \left[ \mathbf{f}_b \mathbf{f}_b^{\mathsf{^{*}T}} \right] = 
       \int_0^{2\pi}  \mathbf{f}_b \mathbf{f}_b^{\mathsf{^{*}T}} D(\theta) d\theta
\end{equation}
where we recall from the introduction that the directional spreading function $D(\theta)$, subject to Eq \ref{eq:5}, is viewed as equivalent to a probability density function. In a diffuse sea, waves from different directions all have the same amplitudes (in an ensemble sense) and as a result $D(\theta)=1/2\pi$. Therefore, in a diffuse sea, the cross spectrum of the blocked moment from Eq (16) can be found as:
\begin{equation} \label{eq:18}
    \begin{split}
        \mathbb{E} \left[ \mathbf{f}_b \mathbf{f}_b^{\mathsf{^{*}T}}\right]_{\text{direct}} 
        & =  
          \int_0^{2\pi}  \mathbf{f}_b \mathbf{f}_b^{\mathsf{^{*}T}} D(\theta) d\theta \\ 
       & =  S_{\eta \eta} (\omega) d\omega \frac{16\rho^2 g^2}{k^6}
        \left(  \frac{kd \sinh (kd) + 1 - \cosh (kd)}{\cosh(kd)} \right)^2 \\
        &  \frac{1}{ (J_1^{'2}(ka) + Y_1^{'2}(ka) )} 
     \begin{bmatrix}
      1/2 & 0 \\
      0 & 1/2
    \end{bmatrix}        
    \end{split}                
\end{equation}
It can be seen that the results from the direct approach, where Eq \ref{eq:18} gives the cross spectrum of the blocked moments by solving the diffraction problem directly, and the reciprocal approach, where Eq \ref{eq:15} gives the cross spectrum of the block moment via DFR approach, are exactly the same, thus verifying the validity of the DFR relationship for offshore applications. 

A full diffraction analysis normally requires computations for wave incidence from all directions. In this section, with an application to a simple articulated buoy, we have demonstrated that by taking advantage of the analytical DFR relation, this $\theta$-wise integration can be avoided using the reciprocal approach instead for a diffuse sea excitation. This would largely improve the computational efficiency and allow a very quick estimation for the cross spectrum of the blocked force and the resulted structural responses. 

\section{Discussions}
\label{sec:4}

\subsection{An efficient reciprocal approach in spreading seas}
\label{sec:4.1}
The response of an offshore structure can be sensitive to both the frequency and the heading of an incident ocean wave, and the most realistic way of ensuring that this effect is captured in a numerical simulation is to model the wave environment as a random spreading sea. 

If a frequency domain analysis is performed, then the dynamic response must be computed at a large number of combinations of frequency and wave heading. This can lead to large computational costs, particularly so at the early design stage when many possible design configurations need to be analysed. According to the design recommendation of structural design of offshore ships from DNV-RP-C102 \cite{DNV2002Recommended}, when a spreading function is applied for the analysis, the wave heading angle spacing should be equal or less than 30 degrees. That is minimum 7 wave angles to be analysed (in case of “all heading included” this results in at least 12 heading angles). Considering the wide design space and large number of design variations, an analysis using a spreading sea is often infeasible during the early design stage, particularly for analysis like fatigue. 

The proposed diffuse field reciprocity (DFR) method provides a new efficient approach to estimate the hydrodynamic response. The improvement of efficiency is a result of the replacement of the integration over wave headings by an ensemble average as seen in Eq \ref{eq:7}. 

Moreover, the analytical reciprocal relationship that relates the wave loading directly to the structure’s radiation behaviour makes DFR principle well positioned for offshore engineering applications. This is because the hydrodynamic radiation problems have been extensively studied and the resulting hydrodynamic coefficients are well documented. 

For simple geometries, the analytical expressions are readily available. For example, a floating circular cylinder in finite-depth water in \cite{Yeung1981Added} and a vertical surface-piercing circular cylinder extending to the seabed and undergoing horizontal oscillations in \cite{Rahman1993Evaluation}. For a large structure with more complicated geometries or flexible structures with multiple modes, the numerical approaches such as boundary element or finite element method can be applied to compute hydrodynamic coefficients. For example, a coupled finite element and boundary element method is used in \cite{Gao2011Hydroelastic} to study a plate-water model, in which the radiated potential is decomposed using modal expansion method in correspondence to structural deflections. Wang and Chen \cite{Wang2015efficient} used higher order boundary element method to solve the added mass and damping coefficients for a FPSO system (floating production, storage and offloading system). Jonkman \cite{Jonkman2010Definition} calculated hydrodynamic added mass and damping matrices for all six rigid body DoFs of the OCE-Hywind spar-buoy using WAMIT. Therefore, the proposed reciprocal DFR method takes advantage of the readily available hydrodynamic coefficients and offers designers and engineers a fast option to estimate the structural response in a random spreading sea.

The DFR approach is based on potential flow assumption and as a result, the drag force contribution has been neglected. In cases that the drag mainly provides damping effect, such as for offshore floating structures of large dimensions, it is expected that DFR can still provide reasonable estimations for the hydrodynamic response because the drag force would be much smaller than the inertia force. And this is demonstrated in Section \ref{sec:5} using a spar type floating wind turbine. 
\subsection{A fast upper bound estimation}
\label{sec:4.2}
Wave spreading functions are generally dependent on the location of the site and the wave height \cite{Forristall1998Worldwide}. When the spreading function is not fixed at early design stages, offshore designers and engineers often adopt a unidirectional sea for their hydrodynamic analysis. 

This was generally thought to be the most conservative design practice as all the wave energy focused in one direction. However, this is not true in all situations because the transfer function $G(\omega,\theta)$ in Eq (1) is also dependent on the direction of excitation. In cases where the structure has motions decoupled from the assumed excitation direction or the sensitive wave incident direction cannot be identified easily, the overall responses can be underestimated. 

On the contrary, all degrees of freedom for the structure are considered at the same time using the proposed diffuse field reciprocity (DFR) approach. If the wave spectrum in the diffuse field is chosen to be the peak amplitude of a spreading sea, an upper bound for the dynamic response can be obtained efficiently. 

To show that, we can take the peak amplitude of a spreading sea: 
\begin{equation} \label{eq:19}
       S_{\eta \eta}(\omega, \theta) = S_{\eta \eta}(\omega) D(\theta) 
       \leq \max_{\theta \in \left[0, 2\pi \right]} 
       \{ D(\theta) \} S_{\eta \eta}(\omega)  = D_0 S_{\eta \eta}(\omega) 
\end{equation}
where $D_0$ is generally smaller than one and commonly used spreading functions can be found in \cite{Goda1999Comparative}. Substitute this peak amplitude for the diffuse field in Eq \ref{eq:6} , the upper bound hydrodynamic response can be obtained as: 
\begin{equation} \label{eq:20}
    \begin{split}
      \bm{\sigma}^2_\zeta & \leq D_0 \int \mathbf{H} (\omega) 
      \left[ \int_0^{2\pi} \mathbf{f}_b (\omega,\theta) \mathbf{f}_b^{\mathsf{^{*}T}}(\omega,\theta) d \theta \right] 
     \mathbf{H}^{\mathsf{^{*}T}}(\omega)S_{\eta\eta} (\omega) d\omega \\
     & = 
      2 \pi D_0 \int \mathbf{H} (\omega) 
      \mathbb{E}_{\theta} \left[ \mathbf{f}_b \mathbf{f}_b^{\mathsf{^{*}T}}  \right] 
     \mathbf{H}^{\mathsf{^{*}T}}(\omega)S_{\eta\eta} (\omega) d\omega
    \end{split}
\end{equation}
and this can be estimated efficiently using the diffuse field reciprocity (DFR) relationship derived in this paper. 

For example, one commonly used model for the spreading function is the cosine-squared model:
\begin{equation} \label{eq:21}
     D(\theta) =
    \begin{cases}
      \frac{2}{\pi} \cos^2(\theta - \theta_p) & \text{for} |\theta -\theta_p| \leq \pi/2\\
      \qquad 0 & \text{otherwise}
    \end{cases} 
\end{equation}
where $\theta_p$ is the dominant wave direction and is assumed to be zero in the following analysis. So the peak amplitude in this case is then given as:
\begin{equation} \label{eq:22}
     D_0 = \max \{ D(\theta)\} = \frac{2}{\pi}
\end{equation}
In a strict sense, scaling the diffuse field by $2\pi D_0$ in Eq \ref{eq:20} violate the constraint given in Eq \ref{eq:5} where the total integration of the spreading function should be one. On the other hand, Eq \ref{eq:20} is equivalent to assume the worst case for wave excitation from independent wave headings ranging from 0 to $2\pi$. Only that with DFR approach, this worst case response can be obtained with a single calculation. When the wave spreading function is unknown, $D_0$ can be taken as one in Eq \ref{eq:22}. This provides a new way to estimate the worst possible responses for the design. In addition, as an upper bound, Eq \ref{eq:20} can be incorporated very efficiently in design optimization iterations to minimize the worst responses.
\section{Demonstration study with a spar type floating wind turbine}
\label{sec:5}
\begin{figure}[!h]
	\centering
	 \subfloat[\centering Hywind spar wind turbine ] {{\includegraphics[width=5cm]{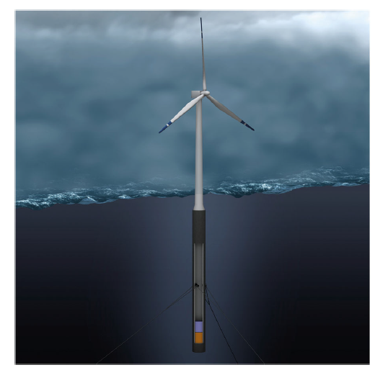} } \label{fig:2a}} 
	 \qquad
    \subfloat[\centering  rigid body model for Hywind]{{\includegraphics[width=5cm]{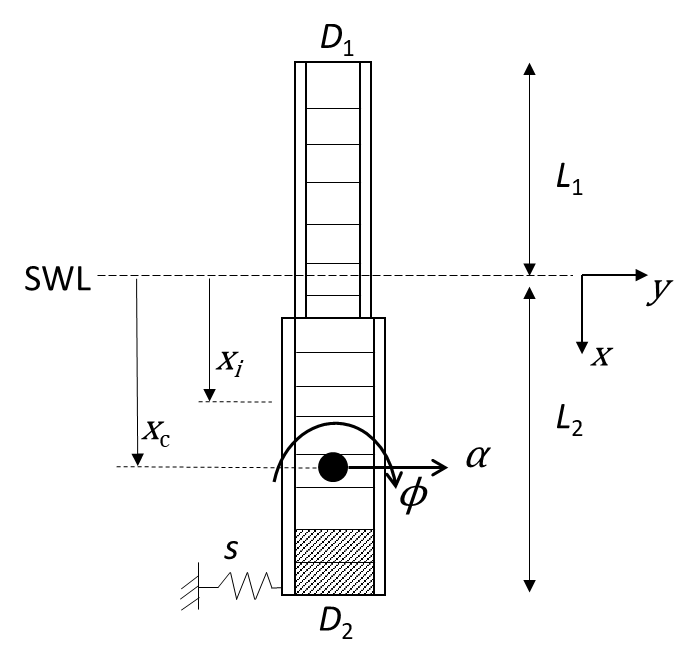} }\label{fig:2b}}
	\caption{A simple rigid body model for Hywind spar wind turbine.}
	\label{fig:2}
\end{figure}

Floating offshore wind turbines (FOWTs) enables economic offshore wind electricity generation in deep waters where fixed foundation turbines are not feasible. Hywind Scotland is the world's first commercial wind farm using floating wind turbines and it consists of a spar type platform \cite{Skaare2015Analysis}. A sketch of the Hywind FOWT is shown in Figure \ref{fig:2a}. In this paper, the Hywind spar type floating platform is used to demonstrate the application of the diffused field reciprocity (DFR) to offshore floating systems.

To apply the DFR method, we need two elements: 1) the hydrodynamic coefficients for the direct field dynamic stiffness matrix. As stated earlier, these coefficients are mostly well documented or ready to be extracted from standard numerical methods \cite{Gao2011Hydroelastic}. This is exactly the case here for the Hywind turbine where we directly extract these coefficients from the literature. 2) the transfer function matrix that relates the structural response to blocked forces. To construct the transfer function, a simplified rigid body model for the FOWT is introduced. These two elements are discussed in details in the next two subsections. And the numerical results are given in Section 5.3.
\subsection{Hydrodynamic added mass and damping matrix}
\label{sec:5.1}
The hydrodynamic added mass and damping for OC3-Hywind spar wind turbine have been computed by Jonkman \cite{Jonkman2010Definition}, where the linear potential flow problem was solved using the WAMIT computer program. WAMIT uses a three-dimensional numerical-panel method in the frequency domain to solve the linearized potential-flow hydrodynamic radiation and diffraction problems for the interaction of surface waves with offshore platforms of arbitrary geometry. 
\begin{figure}[!h]
	\centering
	 \includegraphics[width = 12 cm]{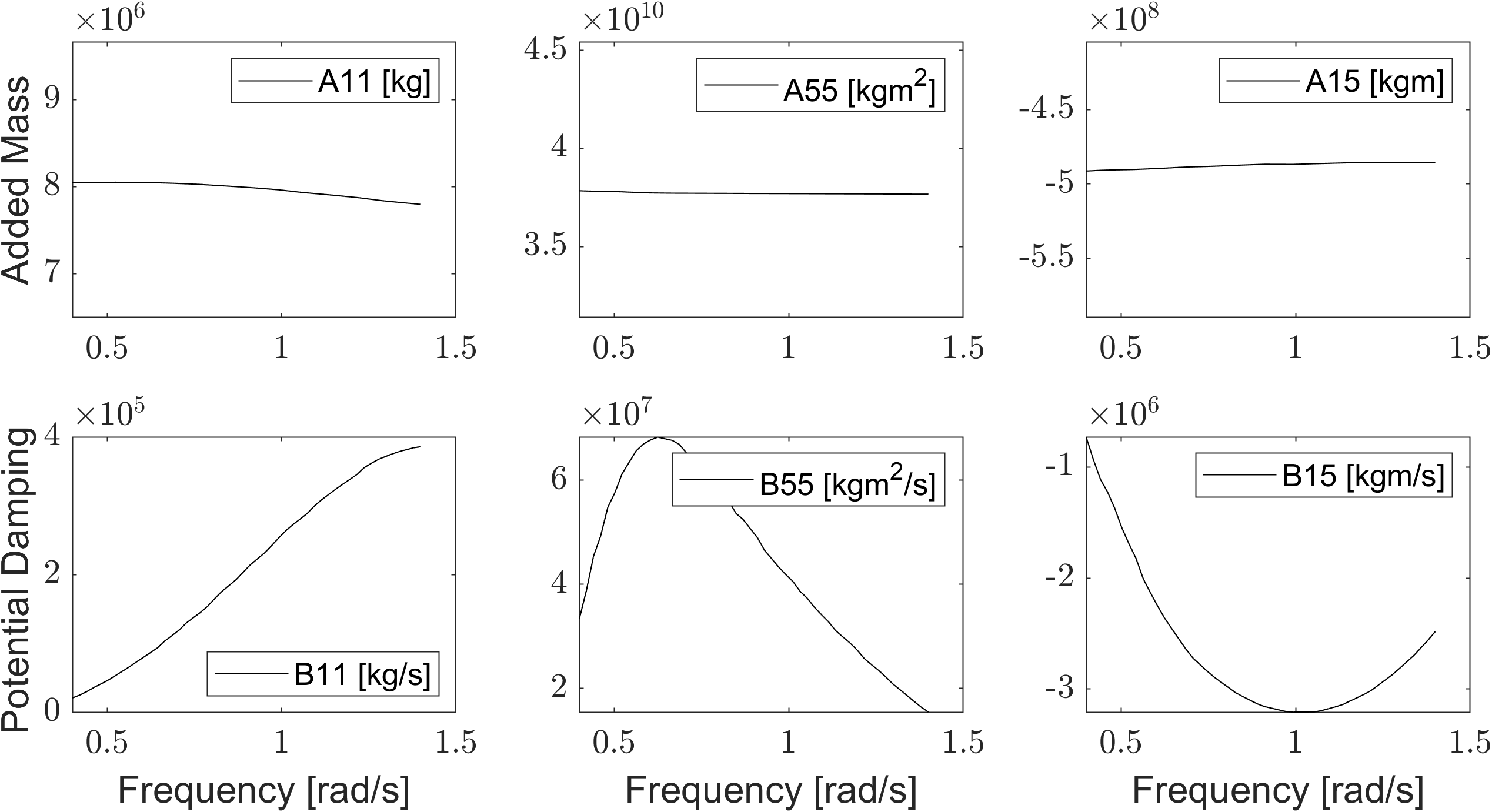}  
	\caption{Hydrodynamic added mass (A11, A55 and A15) and damping (B11, B55 and B15) for OC3-Hywind spar \cite{Jonkman2010Definition}}
	\label{fig:HywindMC}
\end{figure}

These coefficients are reproduced in Figure \ref{fig:HywindMC} for ease of reference. A11, A55 and A15 are the added mass coefficients for surge, pitch and coupled surge-pitch motion respectively, and B11, B55 and B15 are the potential damping coefficients correspondingly. Although in \cite{Jonkman2010Definition}, the added mass and damping matrices were computed for all six degrees of freedom, making use of the axis-symmetry of the structure and for simplicity, only the data for surge and pitch motion are used in this paper.

Given the data in Figure \ref{fig:HywindMC}, the $2 \times 2$ added mass matrix $\mathbf{M}_a$ and potential damping matrix $\mathbf{C}_p$ can then be formed correpondings to the surge and pitch motion. These can then be used in the DFR equation for hydrodymamic response estimation in a random spreading sea. 
\subsection{Transfer function matrix}
\label{sec:5.2}
In frequency domain, we can formulate the transfer function matrix as below:
\begin{equation} \label{eq:23}
       \mathbf{H}(\omega) =  \left[ 
       -\omega^2 (\mathbf{M}_s + \mathbf{M}_a) 
        +i \omega (\mathbf{C}_p + \mathbf{C}_D)  + K 
        \right]^{-1} 
\end{equation}
where $\bf M$, $\bf C$ and $\bf K$ are the inertia, damping and stiffness matrices. The inertia matrix includes the inertia of the structure $\mathbf{M}_s$ and the added mass $\mathbf{M}_a$ due to fluid acceleration. The damping matrix includes both potential damping $\mathbf{C}_p$ and viscous damping $\mathbf{C}_D$. The potential damping is the result of waves travelling away due to radiation in a potential flow, while the viscous damping accounts for effect due to wave separation. As a simplification, the viscous damping part can be approximated using the linearized drag force from Morison’s equation \cite{Low2006Time}. In addition to estimating the viscous damping, as a comparison to the DFR approach, we can also compute the structural response using Morison’s equation with a unidirectional sea state. The details for the formulation of the matrices in Eq \ref{eq:23} are given in Appendix \ref{appendix:c}.
\subsection{Results}
\label{sec:5.3}
\textbf {a) List of parameters}  \\
The structure is approximated as two uniform sections, as shown in Figure 2(b), with constant diameter $D_1$ and $D_2$ respectively.

The values used for the simplified model in this study are listed in Table 1 and they are basically same as those given in \cite{Jonkman2010Definition}. Except that in our case the $D_1$ section has its top level at 6 m below SWL and this gives approximately the same volume as the original structure which has a tapered section between $D_1$ and $D_2$. The simplified model has its centre of mass at 89.7 m and it is very close to the value given in \cite{Jonkman2010Definition} which is 89.9 m. 
\begin{table}
    \footnotesize
	\caption{List of parameters}
	\centering
\makebox [\textwidth][c] 
  {
	\begin{tabular}{*{9}{c}}
		\toprule
			Fluid & \multicolumn{5}{c}{Structure}  & Mooring Line & \multicolumn{2}{c}{Morison's Coefficients}  \\
	      \cmidrule(lr){1-1}   \cmidrule(lr){2-6}  \cmidrule(lr){7-7}  \cmidrule(lr){8-9}   
	        Density 
	        & \multicolumn{1}{p{1cm}}{\centering Density}
	        &  \multicolumn{1}{p{2cm}}{\centering Length above \\ fluid surface  }
	        &  \multicolumn{1}{p{2cm}}{\centering Length below\\ fluid surface  }
	        & Diameter 
	        & \multicolumn{1}{p{1.5cm}}{\centering Centre of \\ mass } 
	        & Stiffness 
	        & Inertia 
	        & Drag \\
		    \cmidrule(lr){1-1}  \cmidrule(lr){2-6} \cmidrule(lr){7-7} \cmidrule(lr){8-9} 
	        $\rho_f$  &  $\rho_s$  & $L_1$  & $L_2$ & $D_1D_2$ & $x_c$ & $s$ & $C_a$ & $C_d$ \\
	           \cmidrule(lr){1-1}  \cmidrule(lr){2-6} \cmidrule(lr){7-7} \cmidrule(lr){8-9} 
	        1.03E+03	&	8.50E+03	& 87	& 120 &	 6.5/9.4 & 	89.7	&	3.80E+09	& 1 & 	1 \\
	       kg/m$^3$ &	 kg/m$^3$	&  m	& m	& m	& m	&	N/m	& -   &	-\\
		\bottomrule
	\end{tabular}
	}
	\label{tab:1}
\end{table}

\textbf{b)	 Wave spectrum} \\
A JONSWAP wave spectrum, with wave significant height of 2.5 m and peak period of 0.8 rad/s, is used here. When using the reciprocal DFR approach, the wave spectrum $S_{\eta \eta}(\omega)$ can be used directly in Eq \ref{eq:10} to get the blocked forces. 

When Morison’s method is used, assuming the area under the spectrum segment is equal to the variance of the wave component, the wave component amplitude is then related to the random wave spectrum, i.e., $a_w(\omega) = \sqrt{2S_{\eta \eta}(\omega) d\omega}$. And this can be used in the Morison’s equations to obtain the wave forces. 

\textbf {c) Blocked force in a diffused field} \\
Morison’s equation estimates the wave forces empirically and takes account of the diffraction effect using the coefficient $C_a$ ($C_a=1$ is used here as Hywind is of circular cylinder shape). When the drag force contribution is small, as in this case for spar type FOWT with large diameter, it is expected that the estimation of blocked forces from the DFR approach should agree with the result using Morison’s equation. 

If the blocked forces due to an incident wave at angle $\theta$ is denoted as $\mathbf{f}_M$ (see for example Figure \ref{fig:1b}), its component with respect to the degrees of freedom $\zeta$ can then be resolved as $\mathbf{f}_b =\mathbf{f}_M \cos \theta$. As we are interested in the response in a diffuse field, we can take the ensemble average of the above blocked force from Morison’s equation over wave headings from 0 to $2\pi$:
\begin{equation} \label{eq:24}
    \begin{split}
        \mathbb{E} \left[ \mathbf{f}_b \mathbf{f}_b^{\mathsf{^{*}T}}\right]_{\text{Morison}}
        & =  
          \int_0^{2\pi}  \mathbf{f}_b \mathbf{f}_b^{\mathsf{^{*}T}} D_{\text{diff}}(\theta) d\theta \\ 
        & =  
       \int_0^{2\pi}  (\mathbf{f}_M \cos \theta) (\mathbf{f}_M^{\mathsf{^{*}T}} \cos \theta) \frac{1}{2\pi} d\theta  \\
        & =  
        \frac{1}{2} \mathbf{f}_M  \mathbf{f}_M^{\mathsf{^{*}T}}
    \end{split}                
\end{equation}
where $D_\text{diff}(\theta) = 1/2\pi$ as there is equal probability of waves coming in any directions, as discussed in the introduction. 
\begin{figure}[!h]
	\centering
	 \includegraphics[width = 12cm]{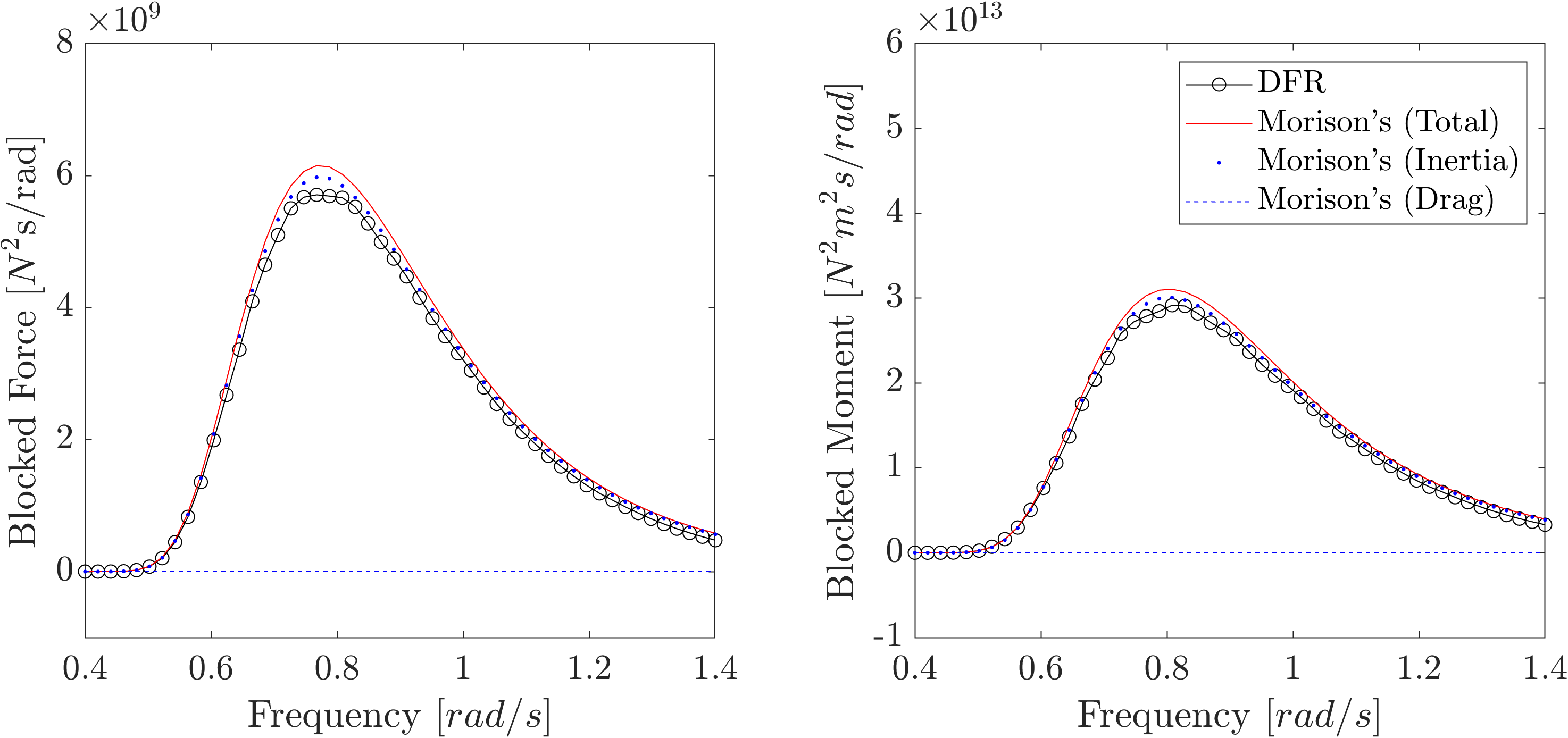}  
	\caption{Ensemble averaged power spectrum of blocked force (left) and moment (right) from diffuse-field reciprocity (DFR) approach and Morison’s equation. Morison’s equation includes both inertial and drag force contributions. The magnitude of the drag contribution is not zero but many orders smaller. }
	\label{fig:DFR1}
\end{figure}
The spectra of the wave forces on the FOWT, from both DFR approach and Morison’s equation, are compared in Figure \ref{fig:DFR1}. As the rigid body model shown in Figure \ref{fig:2b} is an approximation to the Hywind structure, it is not expected that the model based results from Morison’s equation to match exactly the DFR results based on the hydrodynamic coefficients from Hywind. Nevertheless, it is clear from Figure \ref{fig:DFR1} that the two approaches agree very well on the estimation of the wave forces. 

Although the DFR approach is based on potential flow assumption and as a result the drag force contribution has been neglected, for a typical spar type floating wind turbine, it is expected that the linearized drag mainly provides damping effect and the drag force would be much smaller than the inertia force because the structure diameter is large. This is confirmed in Figure \ref{fig:DFR1} where the Morison’s force is dominated by inertia contribution. 

\begin{figure}[!h]
	\centering
	 \includegraphics[width = 12cm]{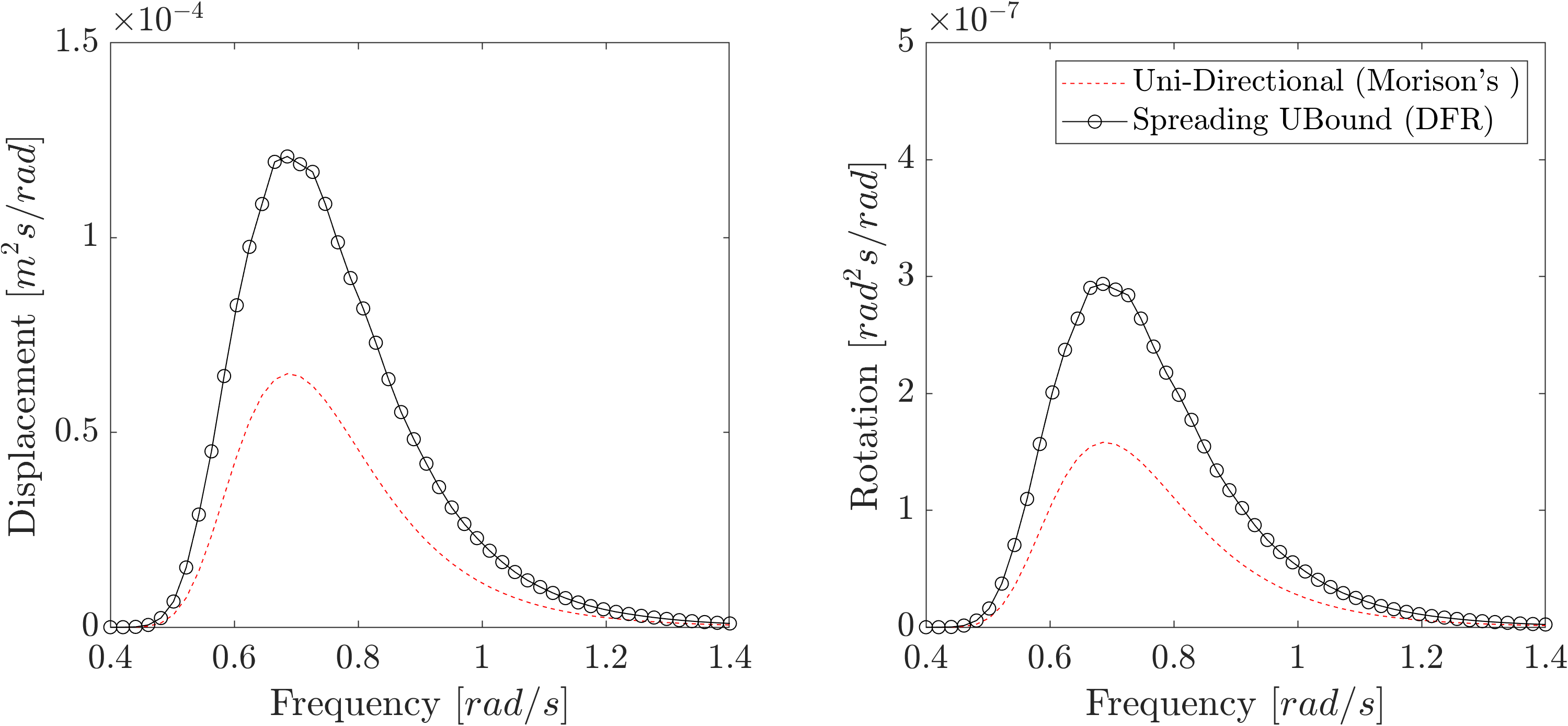}  
	\caption{Spectrum of structural response, rigid body displacement (left) and rotation (right), with the upper bound (UBound) estimated from diffuse-field reciprocity (DFR) approach, in comparison with the results from Morison’s equation.}
	\label{fig:DFR2}
\end{figure}

\textbf{d)	Upper bound of response variance} \\
As discussed in Section 4.2, the proposed DFR approach provides a very efficient way to estimate the worst possible responses for the design of offshore structures. The upper bound of the structural response spectrum with the normalization based on the cosine-squared spreading function in Eq \ref{eq:21}, is shown in Figure \ref{fig:DFR2} for both rigid body displacement and rotation of the FOWT. The results are computed using the upper bound expression in Eq \ref{eq:20}, and the DFR relation derived in Eq \ref{eq:10}. In comparison, the response due to a unidirectional wave using Morison’s equation is also shown in Figure \ref{fig:DFR2}. 

Offshore designers and engineers often adopt a unidirectional sea for their hydrodynamic analysis, because it is simple and the response tends to be on the conservative side. However, in cases where the structure has motions decoupled from the assumed excitation direction or the sensitive wave incident direction cannot be identified easily, the overall responses can be underestimated. 

As can be seen in Figure \ref{fig:DFR2}, using the DFR approach, we can direclty estimate the worst response performance due to a random spreading sea. Although not explicitly demonstrated in this case study with an axisymmetric FOWT, it can be easily seen that, e.g., from Eq \ref{eq:7}, all degree of freedoms of the structure under design can be considered at the same time. Therfore, the proposed reciprocal DFR approach is as efficient as using Morison’s equation for an unidirectional sea excitation, and it can be especially useful for general asymmetric floating systems. 
\section{Conclusions}
\label{sec:6}
A new approach based on the diffuse field reciprocity (DFR) principle has been introduced in this paper to study the hydrodynamic response of offshore systems. This reciprocal approach, first developed for vibroacoustic analysis, is proven to be applicable in offshore context with verification against an analytical hydrodynamic diffraction analysis. 

The DFR approach is efficient because the wave heading integration is avoided in an idealised spreading sea. Moreover, the analytical DFR relationship that relates the wave loading directly to the structure’s radiation behaviour makes DFR principle well positioned for offshore engineering applications. This is because the hydrodynamic radiation problems have been extensively studied and the resulting hydrodynamic coefficients are well documented (e.g., the Hywind example considered in Section \ref{sec:5}). Once the hydrodynamic coefficients are known for the structure under design, the wave loading spectrum is readily available using the reciprocal DFR relationship. 

Using the DFR approach, it is possible to estimate an upper bound response in a spreading sea assuming the worst sea state is coming from all directions. Compared to the commonly used unidirectional sea assumption, the DFR approach provides a new reciprocal approach that is as efficient, considers all degree of freedoms and provides a higher safety factor. And this is demonstrated using a spar type floating wind turbine, where the normalised response from the reciprocal approach is higher than response due to a unidirectional sea. 

Although not explicitly demonstrated in this case study, it is expected that the DFR approach is most useful for general asymmetric offshore systems as all degree of freedoms of the structure under design can be considered at the same time. This would provide offshore designers and engineers a fast tool to estimate the upper bound response of the structure under study and minimize the risk of failure from early design stage. 

\section*{Acknowledgment}
For the purpose of open access, the author has applied a Creative Commons Attribution (CC BY) licence to any Author Accepted Manuscript version arising.
\section*{Data availability statement}
\label{sec:data}
The datasets generated during and/or analysed during the current study are available in the GitHub repository: \url{https://github.com/longitude-jyang/Diffuse-field-reciprocity-for-hydrodynamics}
\bibliographystyle{elsarticle-num} 
\bibliography{references}  

\begin{appendices}
\section{Diffuse field reciprocity relation in offshore context }
\label{appendix:a}

\renewcommand{\theequation}{\thesection.\arabic{equation}}
\setcounter{equation}{0}
The diffuse-field reciprocity (DFR) relation were originally developed in vibro-acoustic field. As a result the parameters in Eq \ref{eq:8} in an offshore context are yet to be defined. In this section, the expressions for the modal density and energy are explained in details. 
\subsection{Modal Density}
\label{sec:a1}
A normal mode of a dynamical system is a pattern of motion in which all parts of the system move sinusoidally with the same frequency and with a fixed phase relation. For a typical system of finite size, distinctive modes can be observed at low frequency, while the high frequency behavior is characterised by a high modal overlap. The level of modal overlap can be quantified by model density $n(\omega)$, which is defined as the number of modes $N$ per radian per second (Fahy, 2007). 

To determine the model density, it is first necessary to find the number of modes. Consider a two dimensional ocean surface of finite area, the number of modes in an area $A$ with wavenumber less than $k$, can be found as shown in Figure 3. The area is shown in wavenumber domain and the wavenumber coordinates are linked with spatial dimensions: $k_1 = l\Delta k_1 = l\pi/l_1$ and $k_2 = l\Delta k_2 = l\pi/l_2$ with $l, m = 1, 2, 3 \dots$. 

The number of modes within the circle can then be found as:
\begin{equation} \label{eq:a1}
     N = \frac{\pi k^2}{4} \frac{1}{\Delta k_1 \Delta k_2} = \frac{Ak^2}{4\pi}
\end{equation}
where $k^2 = k_1^2 + k_2^2$. 

Thus, the modal density in this case can found as:
\begin{equation} \label{eq:a2}
      n(\omega) = \frac{\partial N}{\partial \omega} 
      = \frac{kA}{2\pi}\frac{\partial k}{\partial \omega} 
\end{equation}
From linear wave theory, the dispersion relation for waves in the water is (C. A. Brebbia, 1979):
\begin{equation} \label{eq:a3}
       k\tanh(kd) = \omega^2/g
\end{equation}
where $d$ is the water depth and $g$ is the gravity acceleration constant. 

By differentiating Eq \ref{eq:12}, the modal density per unit $d\omega$ can then be found as below for offshore applications:
\begin{equation} \label{eq:a4}
       n(\omega)  = \frac{k\omega A}{\pi g} \frac{1}{\tanh(kd) + kd \sech(kd)}
\end{equation}
where for the case of deep water,  the modal density $n \approx k\omega A/\pi g$ which clearly increases with frequency and the area as expected. 

\subsection{Energy}
\label{sec:a2}
Consider a single sinusoidal wave of amplitude $a_w$ propagating along the $x$ direction, the particle velocity of the fluid motion are (Newman, 2018):
\begin{subequations} \label{eq:a5}
    \begin{align}
     u = \frac{ga_wk}{\omega} \frac{\cosh(kz - kd)}{\cosh(kd)} cos(kx - \omega t)\label{eq:14a}   \\
     v = \frac{ga_wk}{\omega} \frac{\sinh(kz - kd)}{\cosh(kd)} sin(kx - \omega t )   
     \label{eq:14b} 
    \end{align}
\end{subequations}
Considering a water column of area $A$ and depth $d$, the kinetic energy can then be found as (McCormick, 1973):
\begin{equation} \label{eq:a6}
       T = \frac{1}{2} A \int_d^{-\eta} \rho (u^2 + v^2) d z =  \frac{1}{4} \rho A g a_w^2
\end{equation}
and the total energy can then found as twice the kinetic energy, $E = 2T$, for this sinusoidal wave. 

In the case of a random sea, which is of interest here, the wave elevation is often defined using the one-sided power spectra density, where for a unidirectional sea $S_{\eta \eta} (\omega) d\omega = a_w^2$. Thus, the energy per unit $d\omega$ is then:
\begin{equation} \label{eq:a7}
       E (\omega) = \frac{1}{2} \rho A g S_{\eta \eta} (\omega)
\end{equation}
\section{Analytical derivation of the radiated velocity potential}
\label{appendix:b}

\renewcommand{\theequation}{\thesection.\arabic{equation}}
\setcounter{equation}{0}
To find the velocity potential in the radiated field of the simple articulated buoy, we start with the Laplace’s equation in cylindrical coordinates: 
\begin{equation} \label{eq:b1}
    \nabla^2\phi = \frac{1}{r} \frac{\partial}{\partial r} \left( r  \frac{\partial \phi}{\partial r}  \right)   +
    \frac{1}{r^2} \frac{\partial^2 \phi}{\partial r^2}    +
    \frac{\partial^2 \phi}{\partial z^2}   = 0 
\end{equation}
where $z$ is the coordinate relative to the still water level, as seen in Figure \ref{fig:1a}. 

Making use of the variable separation approach, $\phi(r,\theta,z) = R(r)T(\theta)Z(z)$, the following ordinary differential equations can be obtained:
\begin{equation} \label{eq:b2}
    \begin{split}
          \frac{d^2 Z}{d^2 z}   - k^2 Z & = 0 \\
          \frac{d^2 T}{d^2 \theta}   + m^2 T & = 0   \\
          r^2 \frac{d^2 R}{d^2 r}   + r \frac{dR}{dr} +  R(k^2r^2 - m^2) & = 0
   \end{split}
\end{equation}
where $m$ and $k$ are constants. The last line of Eq \ref{eq:b2} is the Bessel's differential equation, with Bessel functions as the solutions. 

In addition, we have the following boundary conditions:
\begin{equation} \label{eq:b3}
    \begin{split}
          \omega^2 \phi - g \frac{\partial \phi}{\partial z} & = 0  \quad   \text{at} \quad  z=0 \\
          \frac{\partial \phi}{\partial z} & = 0  \quad    \text{at}  \quad  z = -d \\
          \frac{\partial \phi}{\partial z} \Bigr| _{r=a}  & =  -i\omega(z+d) \zeta \cos(\theta) 
   \end{split}
\end{equation}
where the first one is the linearised free surface boundary condition, while the second and third line corresponds to the kinematic boundary condition at the seabed and on the body surface respectively. 

From the boundary conditions, the solutions to the first two ordinary differential equations in Eq \ref{eq:b2} can be found as: $Z(z) = A\cosh[k(z+d)]/\cosh(kd)$ and $T(\theta) = B\cos(\theta)$ with $m=1$ for rigid body motion considered here. $A$ and $B$ are arbitrary constants. 

The solution to the function $R(\cdot)$ in Eq (S19) is the linear combinations of Hankel functions of the first and second kind. For radiated potential, the Sommerfeld radiation condition should be applied. As a result, the second kind Hankel function, which describes inward propagating waves from far field, can be discarded. We then end up with $R(r)= C H^{(1)}_1(kr)$, where $H^{(1)}_q$ is the Hankel function of first kind and $qth$ order. 

Putting the solutions to $Z(r)$, $T(\theta)$ and $R(r)$ together, the radiated velocity potential can then be found as:
\begin{equation} \label{eq:b4}  
\phi(r,\theta,z) =    \sum_{p=0}^{\infty}  B_p
     \frac{\cosh[k_p(z+d)]}{\cosh(k_pd)} 
     H^{(1)}_1(k_pr) \cos(\theta)
\end{equation}
where $k_p$ is the wavenumber from the dispersion relation. The unknown coefficients $B_p$ can be found using the body surface boundary condition in Eq \ref{eq:b3}:
\begin{equation} \label{eq:b5}  
B_p = -i\omega \zeta \frac
{\frac{d}{k_p} \sinh(k_pd) - \frac{1}{k_p^2} [\cosh(k_pd) - 1]}
{k_p \frac{\left[ H^{(1)}_1 (k_p a) \right]'}{\cosh(k_pd)} 
\left[  \frac{\sinh(2k_pd)}{4k_p} + \frac{d}{2} \right]}
\end{equation}
where the derivative of the Hankel function of first kind is $\left[ H^{(1)}_1 (z) \right]' = H^{(1)}_1 (z) / z - H^{(1)}_2(z)$. 
\section{Dynamic analysis for the spar type floating wind turbine}
\label{appendix:c}

\renewcommand{\theequation}{\thesection.\arabic{equation}}
\setcounter{equation}{0}
To obtain the structural mass and stiffness matrix, we model the wind turbine tower as a rigid body structure with two degrees of freedom, translation $\alpha$ and rotation $\phi$, as shown in Figure \ref{fig:2b}. To take account of the non-uniform diameter, the structure is divided into strips and the equation of motion is formed using Lagrange’s equation. For simplicity, the mooring line is simplified as a linear spring. The shaded section at the bottom of structure is of higher structural mass to represent the ballast normally used in spar type FOWT for floating stability. 

The kinetic and potential energy of this rigid body floating tower an be obtained as:
\begin{equation} \label{eq:c1}
    \begin{split}
        T & = \frac{1}{2} \sum_{i=1}^N M_{si}  
        \left[  \dot{\alpha}^2 + (x_i - x_c)^2 \dot{\phi}^2  \right] \\
       V & = \sum_{i=1}^N (M_{si} - B_i)g (x_i - x_c) (1- cos\phi) +
       \frac{1}{2}   \sum_{i=1}^N  s \left[ (x_s - x_c) \phi - \alpha \right]^2
   \end{split}
\end{equation}
where $M_s$ is the structural mass, $B$ is the buoyancy and $s$ is the stiffness of the mooring spring. The subscript $i$ indicates $i_{th}$ strip of the structure. $x$ is the coordinate from the still water level (SWL) and $x_c$ is the coordinate for the centre of mass. 

Applying the Lagrange’s equation and assuming small motion, the equations of motion can be obtained as:
\begin{equation} \label{eq:c2}
    \begin{split}
        & \sum_{i=1}^N M_{si} \ddot{\alpha} - 
       s \left[  (x_s - x_c) \phi - \alpha  \right]  = Q_{\alpha} \\
       &  \sum_{i=1}^N M_{si}  (x_i - x_c) \ddot{\phi} +
     \left[ \sum_{i=1}^N  (M_{si} - B_i)g(x_i - x_c) + \frac{1}{2} s(x_s - x_c) ^2 \right] \phi - s(x_s - x_c) \alpha  = Q_{\phi}
   \end{split}
\end{equation}

Therefore, we could form the mass and stiffness matrix for the structure as: 
\begin{equation} \label{eq:c3}
 \begin{split}
   \mathbf{M}_s  & = 
    \begin{bmatrix}    
         \sum M_{si }  &  0 \\
         0 &  \sum (x_i - x_c)^2 M_{s_i}
   \end{bmatrix}
  \\
     \mathbf{K}  & = 
    \begin{bmatrix}    
         s  &  -(x_s - x_c)s \\
         -(x_s - x_c)s &  \sum (x_i - x_c)(M_{s_i} - B_i)g + (x_s - x_c)^2 s
   \end{bmatrix}    
   \end{split}
\end{equation}

The added mass matrix $\mathbf{M}_a$ and potential damping matrix $\mathbf{C}_p$ for the studied structure are available from literature \cite{Jonkman2010Definition}. Note that the added mass and potential damping were computed with respect to the still water level in \cite{Jonkman2010Definition}. To use it in the transfer function matrix in this case where the degrees of freedom are the translation and rotation of the centre of mass, we need to transform the matrix from the water level to the centre of mass:
\begin{equation} \label{eq:c4}
 \begin{split}
    \hat{\mathbf{M}}_a & = \mathbf{T}_c^{\T} \mathbf{M}_a  \mathbf{T}_c \\
    \hat{\mathbf{C}}_p & = \mathbf{T}_c^{\T} \mathbf{C}_p  \mathbf{T}_c 
   \end{split}
\end{equation}
where $\mathbf{T}_c$ is the coordinate transformation matrix and is given as:
\begin{equation} \label{eq:c5}
   \mathbf{T}_c  = 
    \begin{bmatrix}    
         1  &  x_c\\
         0 &  1
   \end{bmatrix}
\end{equation}

In addition to the potential damping, there is a also damping due to the drag force and this can be estimated as:
\begin{equation} \label{eq:c6}
   \mathbf{C}_D  = 
    \begin{bmatrix}    
         \sum \beta_i  &  -\sum (x_i - x_c) \beta_i \\
          -\sum (x_i - x_c) \beta_i  &   \sum (x_i - x_c)^2 \beta_i
   \end{bmatrix}
\end{equation}
where $\beta_i = 1/2 \gamma_i \rho_f D_i L_i C_d$ and the values of the parameters are given in Table \ref{tab:1}. $L_i$ is the thickness of each strip and it is set as 1 m in this study.  

$\gamma_i$ is the linearization coefficient that commonly used to linearise the Morison's equation. For random wave considered here, $\gamma_i = \sqrt{8/\pi} \sigma_{u_r}(x_i)$. $u_r$ is the relative velocity between the wave and structure and $\sigma$ is the root mean square level of a random signal which can be estimated from its spectrum.

\end{appendices}



\end{document}